\documentclass{svproc}
\usepackage{hyperref}
\usepackage{mathrsfs}
\usepackage{booktabs}   
\usepackage{siunitx}    
\usepackage{url}
\usepackage{graphicx}

\begin{document}
\mainmatter              
\title{Bitcoin Gold, Litecoin Silver: An Introduction to Cryptocurrency's Valuation and Trading Strategy}
\titlerunning{Bitcoin Gold, Litecoin Silver}  
%
\author{Haoyang Yu\inst{1,2} \and Yutong Sun\inst{3} \and Yulin Liu\inst{*4} \and Luyao Zhang\inst{*1}\thanks{ The corresponding author Luyao Zhang is supported by the National Science Foundation China on the project entitled “Trust Mechanism Design on Blockchain: An Interdisciplinary Approach of Game Theory, Reinforcement Learning, and Human-AI Interactions.” (Grant No. 12201266). Haoyang Yu and Yutong Sun are supported by the Summer Research Scholar (SRS) program 2022 under Prof. Luyao Zhang's project entitled "Trust Mechanism Design: Blockchain for Social Good" at Duke Kunshan University.  Haoyang Yu and Yutong Sun completed the research as research assistants of Prof. Luyao Zhang at Duke Kunshan University. Haoyang Yu and Yutong Sun are also the signature work mentees of Prof. Luyao Zhang at Duke Kunshan University. Haoyang Yu, Yutong Sun, and Luyao Zhang are also with SciEcon CIC, a not-for-profit organization aiming at cultivating interdisciplinary research of both profound insights and practical impacts in the United Kingdom. Yulin Liu is also with Shiku Foundation and Bochsler Finance, Switzerland. We thank the anonymous referees at the Future of Information and Communication Conference (FICC) 2024 for their professional and thoughtful comments.}}
\authorrunning{H. Yu, Y. Sun, Y. Liu, L. Zhang} 
%
\tocauthor{}
\institute{
Social Science Division and Data Science Research Center, Duke Kunshan University, Suzhou Jiangsu 215316, China,
\and
Division of Natural and Applied Sciences, Duke Kunshan University, Suzhou Jiangsu 215316, China,\\
\email{*corresponding author:lz183@duke.edu }
\and
Pratt School of Engineering, Duke University, Durham, NC 27708, USA, \\
\and
SciEcon CIC, London, United Kingdom  WC2H 9JQ \\
\email{*corresponding author:yulinzurich@gmail.com}
}

\maketitle              
\begin{abstract}
Historically, gold and silver have played distinct roles in traditional monetary systems. While gold has primarily been revered as a superior store of value, prompting individuals to hoard it, silver has commonly been used as a medium of exchange. As the financial world evolves, the emergence of cryptocurrencies has introduced a new paradigm of value and exchange. However, the store-of-value characteristic of these digital assets remains largely uncharted. Charlie Lee, the founder of Litecoin, once likened Bitcoin to gold and Litecoin to silver. To validate this analogy, our study employs several metrics, including \textsl{unspent transaction outputs (UTXO)}, \textsl{spent transaction outputs (STXO)}, \textsl{Weighted Average Lifespan (WAL)}, \textsl{CoinDaysDestroyed (CDD)}, and public on-chain transaction data. Furthermore, we've devised trading strategies centered around the \textsl{Price-to-Utility (PU)} ratio, offering a fresh perspective on crypto-asset valuation beyond traditional utilities. Our back-testing results not only display trading indicators for both Bitcoin and Litecoin but also substantiate Lee's metaphor, underscoring Bitcoin's superior store-of-value proposition relative to Litecoin. We anticipate that our findings will drive further exploration into the valuation of crypto assets. For enhanced transparency and to promote future research, we've made our datasets available on Harvard Dataverse and shared our Python code on GitHub as open source. 
\end{abstract}

\section{Introduction}
During the Middle Ages, gold and silver played pivotal roles in a multi-currency system, each serving a distinct function. Gold predominantly acted as a reserve, underpinning the value of fiduciary money, whereas silver, with its inherent commodity value, emerged as the frequently chosen medium of exchange \cite{BAUR2018177}\cite{wolters2003silver}. This dichotomy finds a modern echo in the cryptocurrency realm. Charlie Lee, the brains behind Litecoin, once articulated that Litecoin was not an adversary of Bitcoin but rather a complement, envisioned to bridge the functional gaps, specifically as a payment instrument \cite{fortune}. Emulating the reserve-like characteristics of gold, Bitcoin exhibits unique return properties, evidenced by the fact that a significant segment, approximately one-third of Bitcoin investors, is solely on the receiving end without ever initiating transfers or sales \cite{BAUR2018177}. In contrast, echoing silver's agility, Litecoin was architectured for brisker, smaller transactions, boasting a transaction velocity that quadruples Bitcoin's \cite{bhosale2018volatility}. The parallels are hard to miss; Bitcoin and Litecoin mirror the roles gold and silver once held.

Over the past ten years, the cryptocurrency market cap has experienced a meteoric ascent, skyrocketing from humble inception to a staggering 1 trillion US dollars \cite{harvey2021defi}\cite{hardle2020understanding}\cite{halaburda2022microeconomics}\cite{haeringer2018bitcoin}. Yet, their utility transcends mere currency; cryptocurrencies today wear multiple hats, from being products, payment platforms, to securities \cite{cong2021categories}. In real-world scenarios, investors are diversifying portfolios with crypto, and retailers are increasingly integrating it into their payment ecosystems \cite{liu2022cryptocurrency}.

This article casts a spotlight on the "store of value" facet of cryptocurrencies. Anchored in classic monetary tenets, currencies are generally recognized for three cardinal functions: medium-of-exchange, store-of-value, and unit-of-account \cite{liu2022cryptocurrency}. Our narrative zeroes in on the relative positioning of Bitcoin and Litecoin within the cryptocurrency sphere. Historically, gold's intrinsic value led individuals to hoard it, while silver's versatility made it the preferred medium of exchange. Drawing on this, Charlie Lee postulated that Litecoin is to silver what Bitcoin is to gold. But, is this analogy empirically sound? Does Bitcoin unequivocally overshadow Litecoin as a store of value? Is the Bitcoin-Litecoin dynamic truly reflective of the gold-silver equation?

Our methodology leverages public on-chain transactional data and implements measures like \textsl{unspent transaction outputs (UTXO)}, \textsl{spent transaction outputs (STXO)}, \textsl{Weighted Average Lifespan (WAL)}, and \textsl{CoinDaysDestroyed (CDD)} to draw contrasts between Bitcoin and Litecoin. In Section~\ref{2}, we delve into established store-of-value measures and introduce the \textsl{Price-to-Utility (PU)} ratio—a novel cryptocurrency valuation technique centered around the store-of-value principle. Section~\ref{3} unravels our data and findings, while Section~\ref{4} propels the discussion towards prospective research avenues.

\section{The Measures for the Store of Value}
\label{2}
In this section, we delve into established measures pertinent to cryptocurrency valuation, grounded in the concept of store-of-value. These include \textsl{UTXO}, \textsl{STXO}, \textsl{WAL}, \textsl{CDD}, and the \textsl{PU} ratio. While each of these metrics illuminates users' spending behaviors from varied angles, they collectively underscore aspects of the store of value in the realm of cryptocurrencies.

\subsection{Unspent Transaction Outputs (UTXO)}

Blockchain transactions, cataloged in Table~\ref{tab:glossary}, govern the transfer of rights to use cryptocurrencies. Central to this mechanism are the concepts of UTXO (unspent transaction output) and STXO (spent transaction), as outlined by \cite{liu2022deciphering}. Stemming from distributed ledger technology (DLT), the architecture of cryptocurrency transactions has enabled the accurate recording of new transactional activities. The UTXO model, an evolution of DLT, is crucial for verifying new, decentralized cryptocurrency transactions. Notably, UTXOs comprise both block rewards and transaction outputs, with the unique ability to trace every UTXO back to its initial block reward. These block rewards act as incentives, rewarding miners for their commitment to network maintenance \cite{liu2022deciphering}.

\begin{figure}[h]
  \centering
  \includegraphics[width=\linewidth]{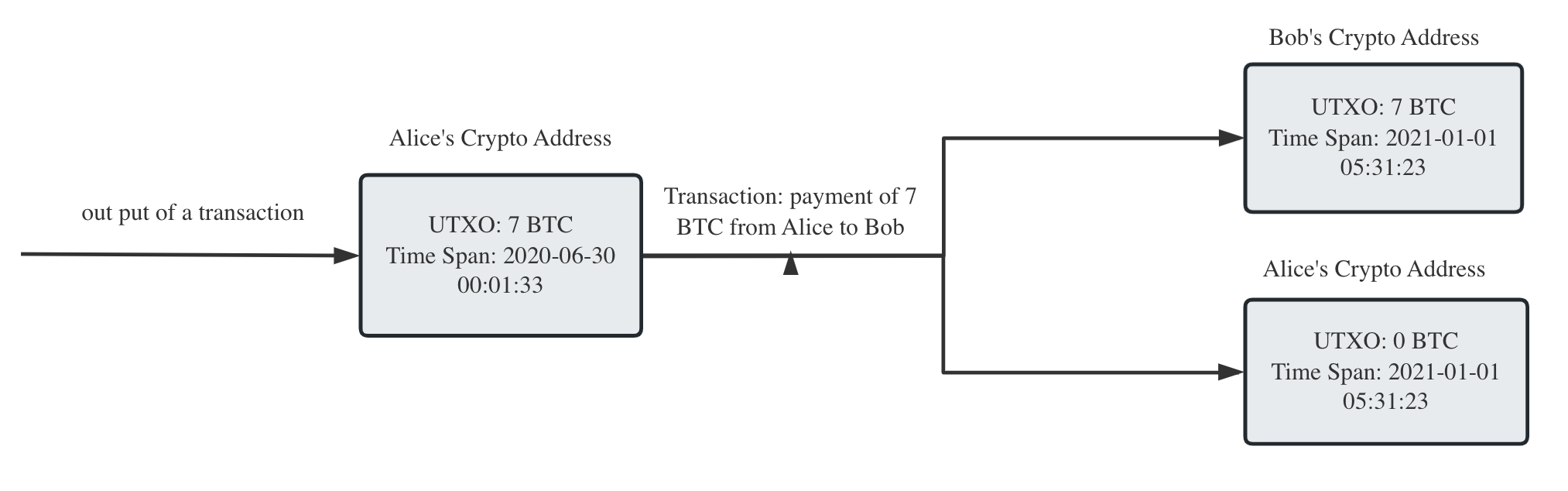}
  \caption{Immediately after Alice’s payment to Bob on January 1, 2021, UTXOs are converted to STXOs with the age of 0.5 years.}
\label{UTXO Transaction}
\end{figure}

For clarity on transaction outputs, envision Alice and Bob, two blockchain users. Alice has 7 unspent Bitcoins, categorized as UTXOs. If she opts to transfer these to Bob, they become STXOs in her ledger. Conversely, when Bob receives the 7 Bitcoins, they are recorded as new UTXOs for him. In this transactional framework, the 7 Bitcoins serve dual roles: as Alice's output and Bob's input. A defining feature of the UTXO blockchain is that each output can be used as an input only once, preserving the uniqueness of each UTXO and STXO. This process is graphically illustrated in Fig.\ref{UTXO Transaction}.
  
\subsection{Spent Transaction Outputs (STXO)}

A UTXO transitions to an STXO once it is used as a transaction input, meaning it has been spent. The term "daily lifespan" is employed specifically for STXOs that were deemed "spent" or "dead" on a given day. This lifespan is calculated by subtracting the time the STXO was generated (or "born") from the time it was spent (or "died") \cite{liu2022deciphering}.

Revisiting the earlier narrative of Alice and Bob can offer more clarity. On the day Alice transfers her 7 Bitcoins to Bob, those Bitcoins are considered "spent" or "dead" for Alice. If she had maintained ownership of these 7 Bitcoins for a duration of 7 years leading up to that day, then they are now characterized as STXOs, each holding a daily lifespan of precisely 7 years.

For a graphical representation, consult Fig.\ref{STXO BTC} and Fig.\ref{STXO LTC}, which showcase the daily lifespan distribution of spent transaction outputs (UTXOs) for both Bitcoin and Litecoin. The duration tokens are held before spending can shed light on the potential storage habits of cryptocurrency users.

\subsection{Weighted Average Lifespan (WAL)}

For a currency to effectively serve as a medium of exchange, it should circulate regularly and not be predominantly hoarded by individuals as a long-term asset \cite{ren2014proof}. One metric to assess this characteristic is the Weighted Average Lifespan (WAL). Instead of simply averaging lifespans, the WAL assigns weights to each unique lifespan based on the number of UTXOs that possess that specific lifespan on a given day. In essence, the WAL represents the average lifespan of UTXOs, with each lifespan weighted by the number of tokens in the respective transaction outputs \cite{liu2022deciphering}. The mathematical representation to compute the WAL is as follows:

\begin{equation}
    WAL[date = i] = \sum_{date=i} (\#_{UTXO} \times Lifespan)/\sum_{date=i} \#_{UTXO}
  \end{equation}

where 
\begin{equation}
    Lifespan = spent\_block\_timestamp-block\_timestamp
\end{equation}

That is, for instance, if three Bitcoins are spent on the day $i$, one has been held as an output for 9 years, and the other two have been held as an output for 6 years, then we can calculate the \textsl{WAL} for the three Bitcoins by:
\begin{equation}
    WAL[date = i] = (9 \times 1 + 6 \times 2) \times (1/3) = 7 years
  \end{equation}

In volatile market conditions, where price fluctuations are significant, the behavior of cryptocurrency transactions can be particularly revealing. Market turmoil often prompts investors to make pivotal decisions about their assets. Those who have invested in cryptocurrencies tend to engage in more frequent transactions during these uncertain times. As a result, the Weighted Average Lifespan (WAL) of tokens in these transaction outputs can serve as a valuable metric to gauge the store of value.

A longer WAL indicates a heightened store of value. It signifies an increased propensity among users to hold onto their assets, suggesting that they anticipate more substantial returns from prolonged holding. In essence, before executing a transaction, users are inclined to hold for extended durations, expecting a more favorable outcome. As depicted in Fig.\ref{WAL BTC} and Fig.\ref{7}, Bitcoin's WAL consistently reaches elevated levels during market upheavals, reinforcing its perceived store of value during tumultuous times.

\subsection{CoinDaysDestroyed (CDD)}

CoinDaysDestroyed (CDD)\footnote{For details, refer to Table~\ref{tab:glossary}. While the concept of CoinDaysDestroyed is discussed in industry documentation, for instance at ~\url{https://academy.glassnode.com/indicators/coin-days-destroyed/cdd-coin-days-destroyed}, it has yet to gain widespread attention in academic literature.} offers insight into both the lifespan and transaction volume of a cryptocurrency. Essentially, it serves as a lens through which spending behavior can be analyzed. Here's how it works: for every day a coin unit remains unspent, it accrues one "coin day". Once that coin is transacted or "destroyed", its accumulated coin days revert to zero, and the counting begins anew. This accumulation is essentially the age of the UTXO, quantified in days. The CDD is calculated using the following formula:

\begin{equation}
    CDD=\#_{UTXO} \times Ages[days]
  \end{equation}

Consider the following scenario: 10 Bitcoins have remained as UTXO for 12 hours (or 0.5 days) since their last transaction. This means the CDD for these 10 Bitcoins amounts to 5 coin days (10 Bitcoins * 0.5 days). Now, let's expand our scope. Within a specific time frame, say a day (represented as date=$i$), various UTXO amounts of a token might exist, each having a distinct age. The formula to calculate the aggregate CDD for these UTXOs on date=$i$ is:

\begin{equation}
    CDD[date = i] = \sum(\#_{UTXO} \times Ages[days])
  \end{equation}

\subsection{Price-to-Utility (PU) Ratio \& Trading Strategy}
The Price-to-Utility (PU) ratio offers a holistic evaluation of a cryptocurrency, encapsulating its utility as the medium of exchange, unit of count, and store of value. Essentially, the PU ratio is determined by dividing the token's price by its inherent utility. The Token Utility (TU) – a measure of the cryptocurrency's utility as a currency – is delineated as follows \cite{liu2022cryptocurrency}:

\begin{equation}
    Token\;Utility = \frac{token\;velocity \times staking\;ratio}{price\;volatility \times dilute\;rate}
  \end{equation}
  
Token velocity acts as a measure of a cryptocurrency's usage as a medium of exchange. Defined mathematically, it represents the percentage of tokens transacted over the past 24 hours compared to the current total token supply. Conversely, the staking ratio gauges the cryptocurrency's function as a store of value, indicating the proportion of tokens with a lifespan exceeding one year. From this perspective, token velocity provides insights into the activity levels of users who utilize the token for exchanges. Meanwhile, the staking ratio sheds light on the confidence long-term holders place in cryptocurrency. Another critical metric is the dilution rate, which signifies the annual growth rate of the token supply. This rate serves as an additional store of value indicator; a higher dilution rate usually corresponds to a diminished store of value, consequently reducing token utility. In terms of the unit of account, the inverse of price volatility is utilized. Given that cryptocurrencies often experience significant price volatility, its inverse helps temper the pronounced token price fluctuations in the PU ratio.

Shifting the focus to trading strategies, the PU ratio-based automated strategy demonstrates superior performance in the cryptocurrency market compared to traditional approaches like buy-and-hold and the moving average (MA) crossover rule \cite{liu2022cryptocurrency}. Specific thresholds have been defined: overvaluation (PU $>$ 100), undervaluation (PU $<$ 60), and a normal range (60 $<$  PU $<$  100). In automated trading terms, a buy signal is triggered when the PU ratio equals or falls below the 0.1 quantiles of the previous day's ratio. Conversely, a sell signal arises when the PU ratio matches or exceeds the 0.9 quantiles of the prior day's historical PU ratio.

\section{Dataset and Results}
\label{3}
\subsection{Data Sources}
\textbf{Data and Code Availability Statements:} For enhanced transparency and to promote future research, we've made our datasets available on Harvard Dataverse\cite{DVN/AFIM6U_2023} and shared our Python code on \url{https://github.com/SciEcon/bitcoin_golden_litecoin_silver} as open source.

Understanding users' spending behavior is pivotal as it offers insights into the primary use cases of a token, be it as a medium of exchange or a store of value. Notably, each token may elicit distinct spending behaviors from its users. While prior research has primarily relied on simulation-based methods to probe spending behavior \cite{campajola2022microvelocity}, our approach diverges by directly quantifying store-of-value metrics. The transaction datasets for Bitcoin and Litecoin, which formed the basis for Figures \ref{2} through \ref{7} and \ref{9} to \ref{13}, were sourced from the BigQuery Public Datasets program\footnote{\url{www.cloud.google.com/blog/products/data-analytics/introducing-six-new-cryptocurrencies-in-bigquery-public-datasets-and-how-to-analyze-them}}. We extracted this data utilizing open-source code by Liu et al. \cite{liu2022deciphering}, titled Deciphering Bitcoin Blockchain Data by Cohort Analysis\footnote{\url{www.nature.com/articles/s41597-022-01254-0}}. These datasets, furnished by Google Cloud, encompass the blockchain transaction histories of multiple cryptocurrencies including Bitcoin, Ethereum, Litecoin, and several others. With the assistance of Python scripts from Liu et al.'s research \cite{liu2022deciphering}, we transformed the raw data into UTXO and STXO formats. Specifically, our queries spanned transaction datasets for Bitcoin from January 3rd, 2009, to May 31st, 2022, and for Litecoin from January 1st, 2011, to May 31st, 2022.

\begin{figure}[hbtp!]
  \centering
  \includegraphics[width=\linewidth]{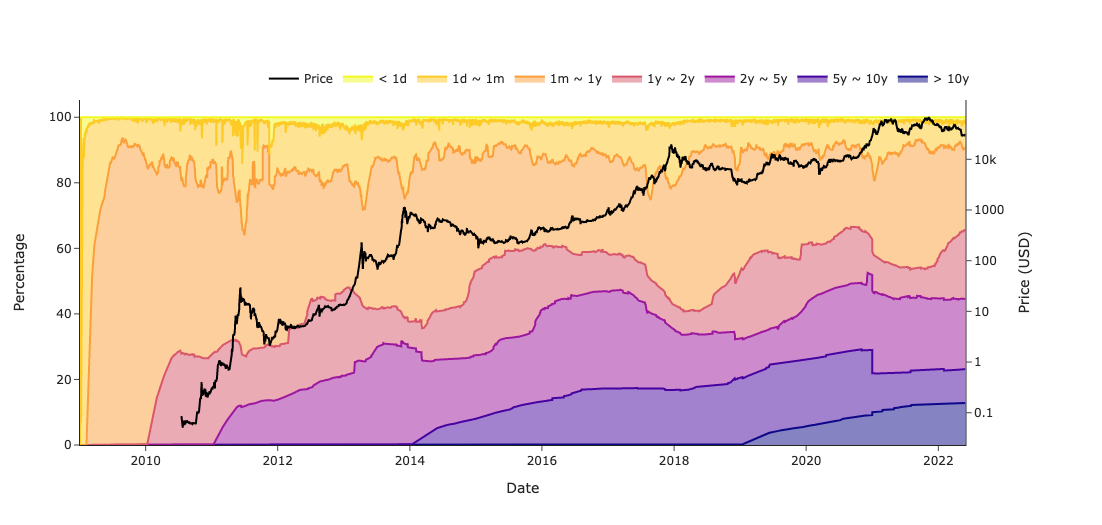}
  \caption{Daily age distribution of Bitcoin UTXOs. It shows the percentage of Bitcoin UTXOs with different ages each day ($<$1day, 1day$-$1month, 1month$-$1year, 1year$-$2years, 2years$-$5years, 5years$-$10years, $>$10years) each day until May 31. 2022.}
\label{UTXO BTC}
\end{figure}

\begin{figure}[hbtp!]
  \centering
  \includegraphics[width=\linewidth]{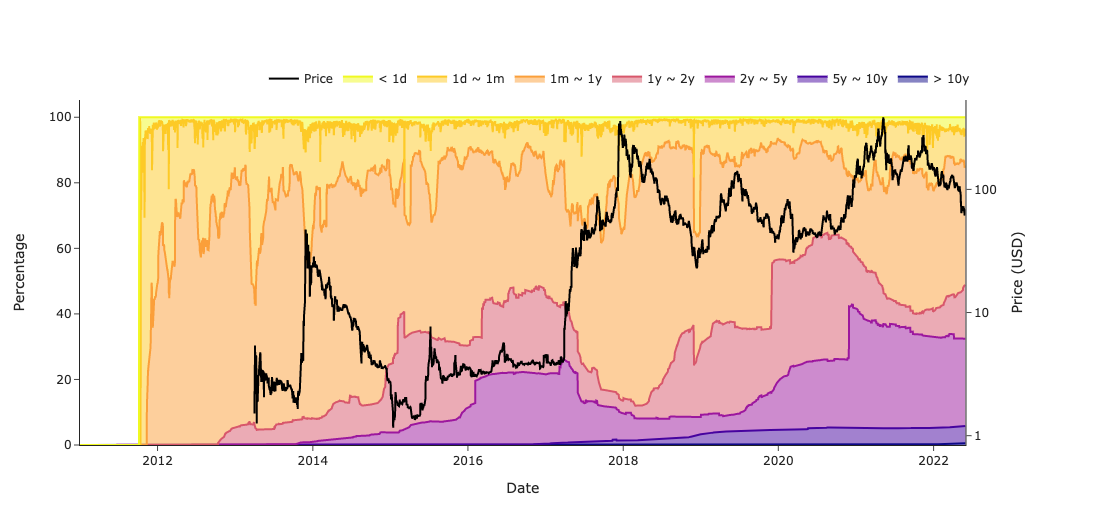}
  \caption{Daily age distribution of Litecoin UTXOs. It shows the percentage of Bitcoin UTXOs with different lifespans ($<$1day, 1day$-$1month, 1month$-$1year, 1year$-$2years, 2years$-$5years, 5years$-$10years, $>$10years) each day until May 31. 2022.}
\label{UTXO LTC}
\end{figure}

\subsection{Results} 
Our findings from UTXO, STXO, and WAL measures uniformly indicate that users typically retain Bitcoin for more extended durations in comparison to Litecoin. Figures \ref{UTXO BTC} and \ref{UTXO LTC} present the visualization results of the daily age distribution of cumulative unspent transaction output (UTXO) for Bitcoin and Litecoin respectively. A deeper dive into Fig.\ref{UTXO BTC} reveals an increasing trend in the ratio of long-age UTXOs (categories: 1 year-2 years, 2 years-5 years, 5 years-10 years, and $>$10 years) for Bitcoin up to May 2022. Conversely, Fig.\ref{UTXO LTC} predominantly highlights the dominance of short-age UTXOs (categories: $<$1 day, 1 day-1 month, and 1 month-1 year) for Litecoin. These patterns suggest a distinct usage paradigm for the two cryptocurrencies: while Litecoin circulates more actively, embodying its role as a medium of exchange, Bitcoin is largely stored, underscoring its status as a valued asset. For a more detailed examination, refer to the appendix. Fig. \ref{STXO BTC} and Fig. \ref{STXO LTC} delineate the results pertaining to STXO, while Fig.\ref{WAL BTC} and Fig.\ref{7} shed light on the WAL findings.

\begin{figure}[hbtp!]
  \centering
  \includegraphics[width=\linewidth]{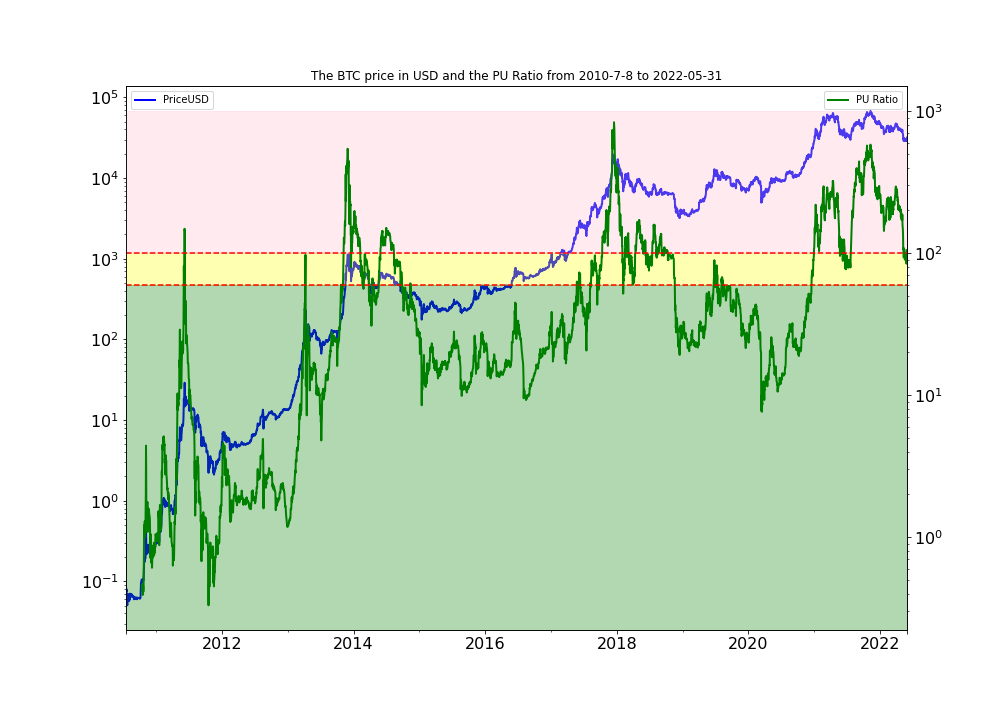}
  \caption{The BTC Price in USD (Blue Line, Left Axis) and PU Ratio (Green Line, Right Axis)}
\label{PURatio1}
\end{figure}

\begin{figure}[hbtp!]
  \centering
  \includegraphics[width=\linewidth]{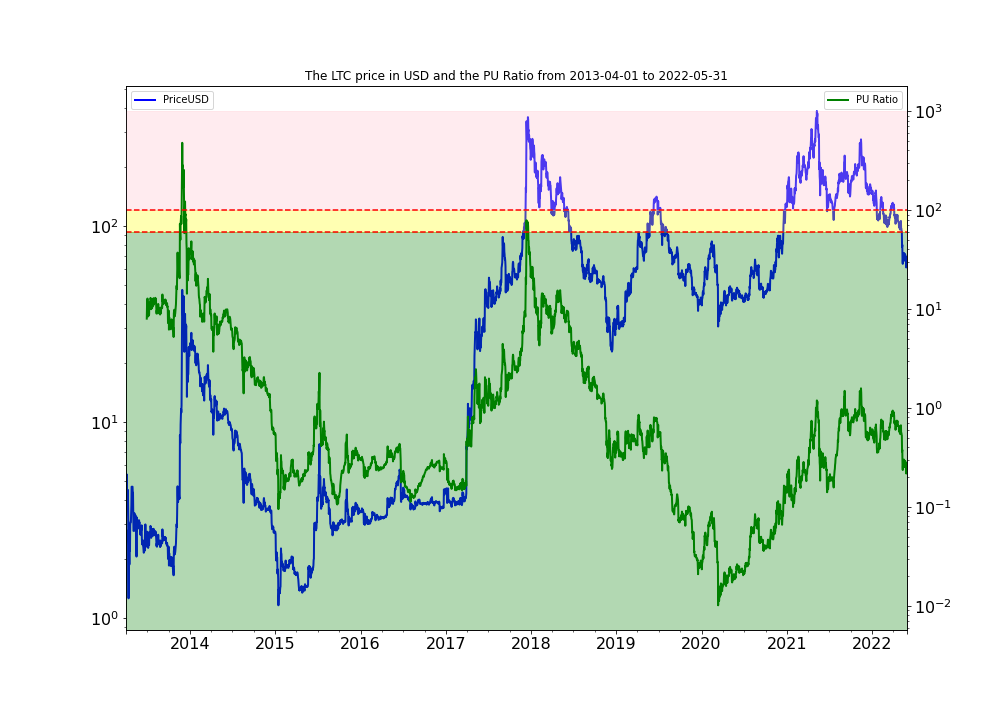}
  \caption{The LTC Price in USD (Blue Line, Left Axis) and PU Ratio (Green Line, Right Axis)}
\label{PURatio2}
\end{figure}

\begin{figure}[hbtp!]
  \centering
  \includegraphics[width=\linewidth]{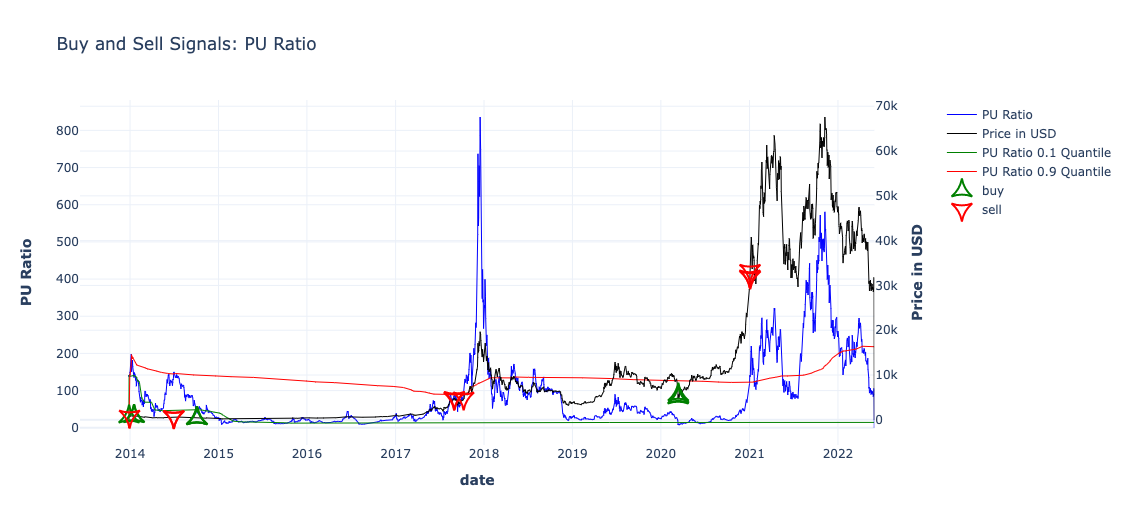}
  \caption{BTC Buy and Sell Signals for the PU Ratio}
\label{In1}
\end{figure}

\begin{figure}[hbtp!]
  \centering
  \includegraphics[width=\linewidth]{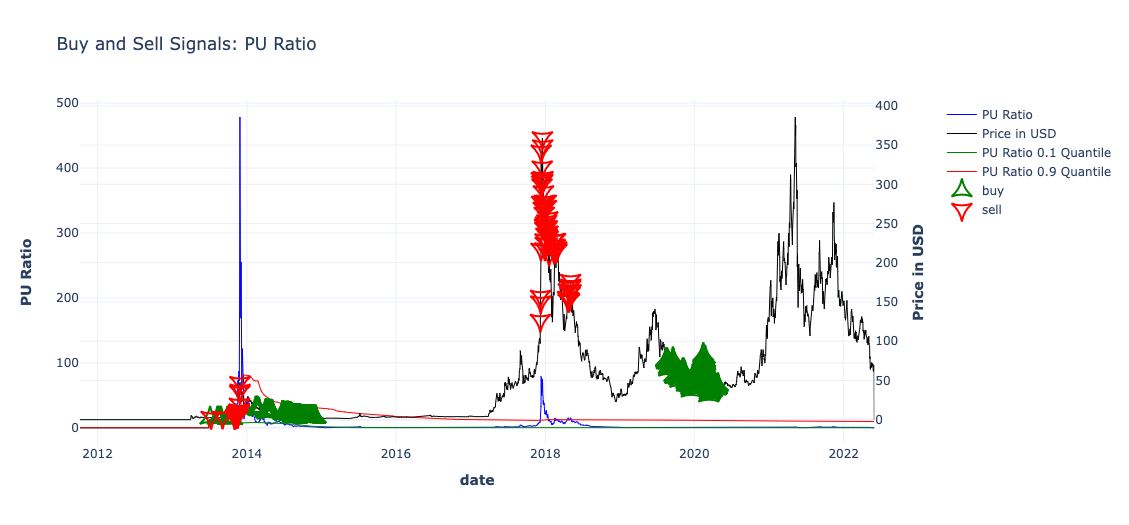}
  \caption{LTC Buy and Sell Signals for the PU Ratio}
\label{13}
\end{figure}

Figures \ref{PURatio1} and \ref{PURatio2} delineate Bitcoin and Litecoin's prices in USD juxtaposed against their respective PU Ratios. Notably:
\begin{itemize}
    \item The red zone (PU $>$ 100) signifies overvaluation.
    \item The yellow zone (60 $<$ PU $<$ 100) indicates a valuation within a normal range.
    \item The green zone (PU $<$ 60) suggests undervaluation.
\end{itemize}
From the visual representation, Bitcoin's PU ratio predominantly hovers within the yellow zone, implying that its price closely aligns with its inherent value. In stark contrast, Litecoin's PU ratio consistently resides in the green zone, marking it as undervalued.

Turning our attention to Fig.\ref{In1} and Fig.\ref{13}, they illustrate the backtesting strategy built on the PU-ratio. Here, a 'buy' signal is triggered when the PU ratio drops to or below its 0.1 quantiles, and a 'sell' signal when the ratio climbs to or exceeds its 0.9 quantiles, derived from historical data. Commencing with an initial capital of 100,000 USD, and incorporating a 0.1\% transaction fee and a transaction cap of 100 units, our analysis encompasses:
\begin{itemize}
    \item Bitcoin's transactions from December 27, 2013, to May 31, 2022.
    \item Litecoin's transactions from October 7, 2011, to May 31, 2022.
\end{itemize}

The culmination of the designated Bitcoin trading period witnessed a robust ROI of 7016.06\% under the PU ratio-based strategy, coupled with an annualized Sharpe ratio of 3.73. In comparison, Litecoin's trading strategy yielded a gross ROI of 2582.20\% and an annualized Sharpe ratio of 2.51. Notably, Fig.\ref{In1} and Fig.\ref{13} underscore a significantly elevated trading frequency for Litecoin when juxtaposed against Bitcoin.

\section{Discussion and Future Work}
\label{4}
Mining Bitcoin demands advanced computer hardware due to its intense processing needs~\cite{pagnotta2022decentralizing}. Harvey et al. \cite{harvey2022investor} juxtapose the operational costs of mining a single Bitcoin with those of extracting an ounce of gold, a pound of copper, and a barrel of oil from 2011 to 2021. On the other hand, Litecoin can be mined with less computational effort, even on standard computers \cite{bhosale2018volatility}.

By assessing different store-of-value metrics, we discern that Bitcoin's role in the cryptocurrency ecosystem mirrors gold in the traditional market, while Litecoin resembles silver. This suggests that Bitcoin primarily serves as a store of value, whereas Litecoin functions more as a medium of exchange.

Such insights prompt intriguing questions for subsequent research. For example, how does a token's valuation correlate with its designated function, as outlined by Cong et al. who classify crypto tokens into categories such as security tokens, utility tokens, and work tokens? \cite{cong2021categories}. Furthermore, how does the performance of the underlying blockchain\cite{zhang2023design,zhang2022cooperation}, a foundational component, influence token utility?

Another intriguing avenue of research would be to delve deeper into the environmental implications of cryptocurrency mining, given the global emphasis on sustainable practices\cite{jiang2021policy}. Moreover, understanding the factors that establish a token's role in the market - whether as a store of value or medium of exchange - could provide insights into market dynamics\cite{zhang2022blockchain,zhang2023understand,ao_are_2022,liu2022,zhang_2022_sok} and sentiments\cite{fu2022ai,Zhang_2023}. Finally, in the realm of practical applications, how might one refine automated trading strategies based on nuanced cryptocurrency valuations\cite{zhang2022data}? These questions, among others, offer exciting directions for future exploration in the cryptocurrency domain.

\bibliographystyle{spmpsci}
\bibliography{reference}

\appendix
\begin{figure}[hbtp!]
  \centering
  \includegraphics[width=\linewidth]{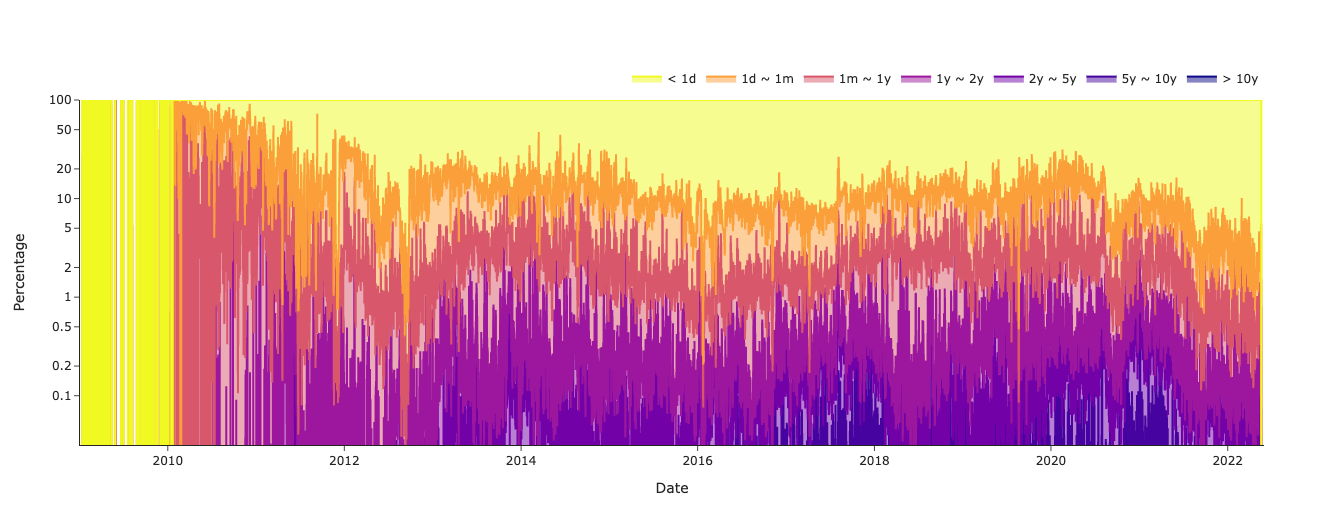}
  \caption{Lifespan distribution of BTC STXOs. The figure shows the log percentage of spent transaction outputs with different lifespans in each day until May 31. 2022. For example, by Feb. 2021, the STXOs with lifespans of less than one day accounted for 80\% of all STXOs, while those with lifespans between 1 day and 1 month accounted for another 15\%. }
  \label{STXO BTC}
\end{figure}

\begin{figure}[hbtp!]
  \centering
  \includegraphics[width=\linewidth]{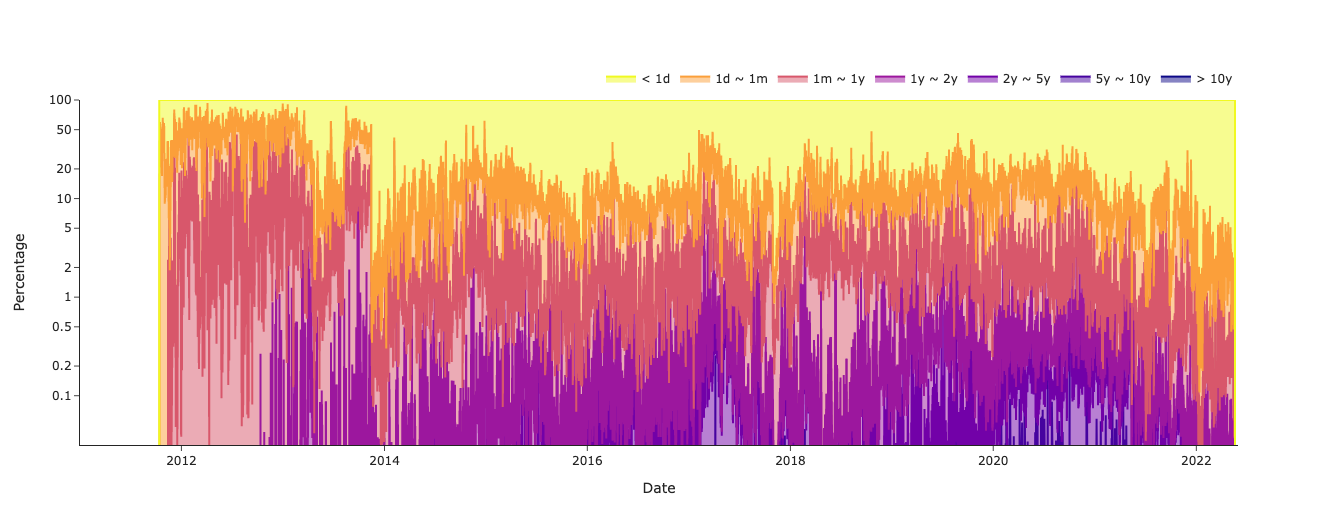}
  \caption{Lifespan distribution of LTC STXOs. The figure shows the log percentage of spent transaction outputs with different lifespans each day until May 31. 2022.}
  \label{STXO LTC}
\end{figure}

\begin{figure}[hbtp!]
  \centering
  \includegraphics[width=\linewidth]{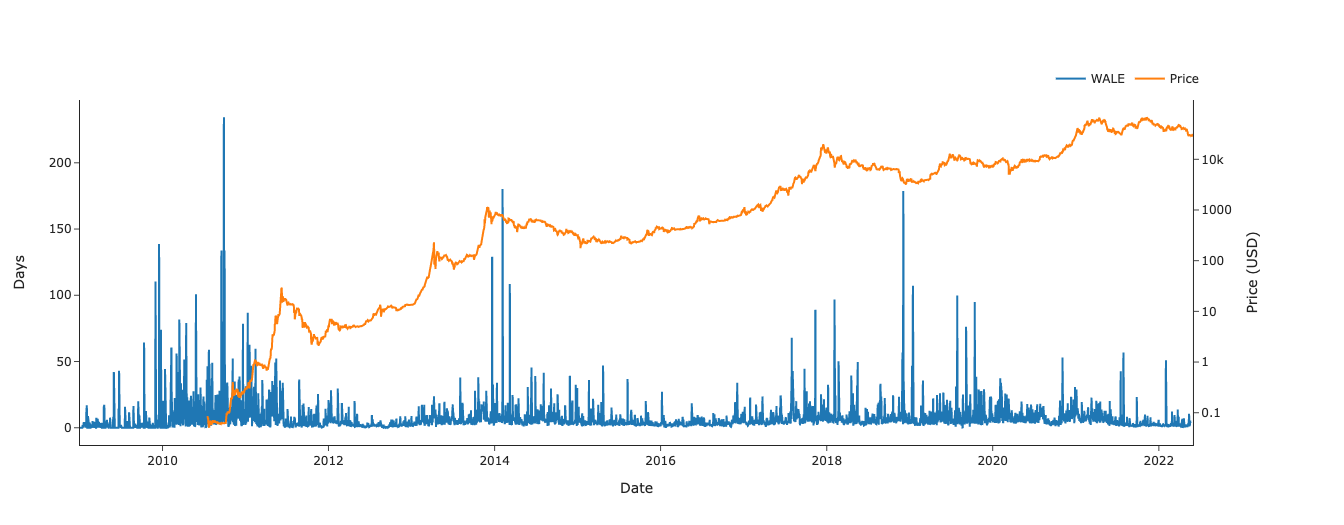}
  \caption{Daily weighted average lifespan(WAL) of Bitcoin UTXOs and BTC price. The figure shows that the WAL of BTCs in UTXOs attains a peak value when the BTC price is volatile. For example, the 2014 peak of WAL value closely followed the rocketing of BTC price from \$100 to \$1000 and its subsequent price collapse. This implies that older BTC becomes more active during market turmoil.}
  \label{WAL BTC}
\end{figure}

\begin{figure}[hbtp!]
  \centering
  \includegraphics[width=\linewidth]{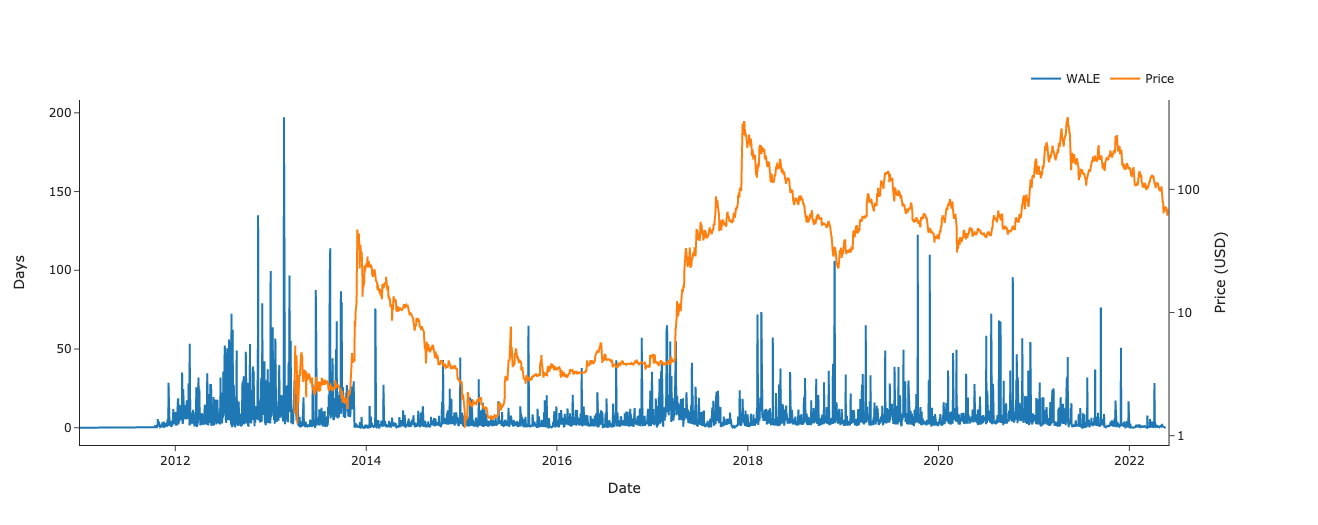}
  \caption{Daily weighted average lifespan(WAL) of Litecoin UTXOs and LTC price. The figure shows that the WAL of LTCs in UTXOs attains a peak value when the LTC price is volatile \cite{liu2022deciphering}.}
  \label{7}
\end{figure}

\begin{figure}[hbtp!]
  \centering
  \includegraphics[width=\linewidth]{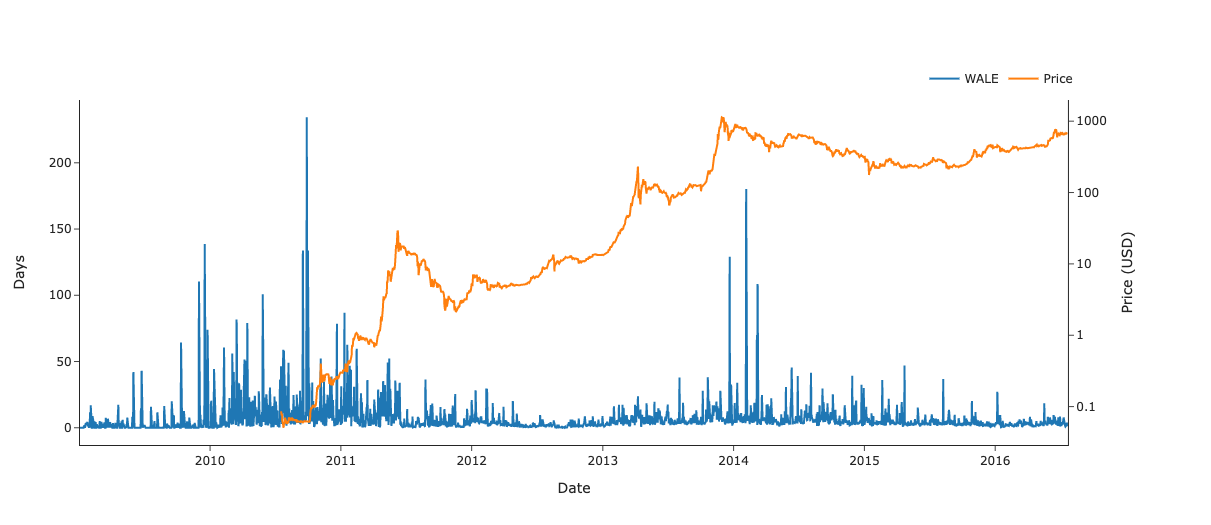}
  \caption{Daily Weighted Average Lifespan of Bitcoin UTXO \& Price from Jan 2009 each day until July 2016.}
\label{9}
\end{figure}

\newpage
\begin{table}
    \caption{Glossary Table}
    \begin{center}
    \begin{tabular}{p{4cm}p{6cm}p{3cm}}
    \toprule
    Terminology & Definition & Source\\
    \midrule
    Blockchain & "A distributed database shared among nodes of a computer network. It maintains a secure and decentralized record of transactions, ensuring data fidelity and generating trust without a central authority." & \href{www.investopedia.com/terms/b/blockchain.asp}{Hayes. 2022.}\\
    UTXO & "The residual digital currency after a cryptocurrency transaction. Represents an unspent transaction output." & \href{www.investopedia.com/terms/u/utxo.asp}{Frankenfield. 2022.}\\
    STXO & "The state of a UTXO when it is used in a transaction. Each UTXO can be spent once, converting it to STXO with a timestamp." & \href{www.arxiv.org/abs/2103.00173}{Liu et al. 2022.} \\
    Fisher's Equation & "Posits a proportionality between monetary mass ($M$) and price level ($P$) via economic output ($Q$) and the velocity of money ($V$)." & \href{www.science.org/doi/10.1126/science.37.959.758.b}{Wilson. 1913.} \\
    WAL\\\tiny Weighted Average Lifespan & "The mean lifespan difference between when a transaction output was spent and created, weighted by the transaction output's BTC amount." & \href{www.arxiv.org/abs/2103.00173}{Liu et al. 2022.} \\
    Coin Days Destroyed & "A metric emphasizing the economic activity of long-unspent coins, offering a deeper insight than raw transaction volumes." & \href{www.academy.glassnode.com/indicators/coin-days-destroyed/cdd-coin-days-destroyed }{Glassnode Academy. 2022.}\\
    Bitcoin & "A digital currency facilitating payments without intermediaries. Miners receive it as a reward for transaction verification. It's also purchasable on various exchanges." & \href{www.investopedia.com/terms/b/bitcoin.asp}{Frankenfield. 2022.}\\
    Litecoin & "A decentralized cryptocurrency based on an altered Bitcoin codebase. It offers reduced transaction fees, swifter confirmations, and expedited mining difficulty adjustments." & \href{www.en.wikipedia.org/wiki/Litecoin}{Wikipedia. 2022.} \\
    Cryptocurrency & "A digital medium of exchange operating on a computer network independent of central control mechanisms, like banks or governments." & \href{www.doi.org/10.5937\%2Fekonomika1801105M}{Milutinović. 2018.} \\
    Tidal Frequency & "The speed, in daily degrees, of a tidal component formed by specific sun-earth-moon system forces." & \href{www.encyclopedia2.thefreedictionary.com/tidal+frequency}{McGraw-Hill Dictionary. 2003.} \\
    Commodity Money & "Objects serving as money due to their intrinsic value aligning with their exchange value." & \href{www.jstor.org/stable/42633661?seq=2#metadata_info_tab_contents}{Wolters. 2003.} \\
    Fiduciary Money & "Token coins or notes whose monetary value exceeds their material value." & \href{www.jstor.org/stable/42633661?seq=2#metadata_info_tab_contents}{Wolters. 2003.} \\
    \bottomrule
  \end{tabular}
  \end{center}
  \label{tab:glossary}
\end{table}

\end{document}